\documentclass[conference,a4paper]{APSIPA2026}
\usepackage{amsmath}
\usepackage{graphicx}
\usepackage{multirow}
\usepackage{threeparttable}
\usepackage{bm,makecell,amsfonts,algorithmic,algorithm}
\usepackage[backend=biber,style=ieee,]{biblatex}
\usepackage{geometry}
\usepackage{fancyhdr}

\fancypagestyle{firststyle}{
  \fancyhf{}
  \fancyhead[C]{2026 Asia Pacific Signal and Information Processing Association Annual Summit and Conference (APSIPA ASC)}
}

\begin{document}

\title{PACodec: A Low-bitrate Neural Speech Codec with Parallel Additive Vector Quantization}

\author{
\authorblockN{
Fei Liu, Yang Ai$^*$, Xiao-Hang Jiang, Zhen-Hua Ling
}

\authorblockA{
National Engineering Research Center of Speech and Language Information Processing,\\ 
University of Science and Technology of China, Hefei, China\\
E-mail: \{fliu215,jiang\_xiaohang\}@mail.ustc.edu.cn, \{yangai,zhling\}@ustc.edu.cn}
}

\maketitle
\thispagestyle{firststyle}
\pagestyle{empty}

\begin{abstract}
This paper proposes PACodec, a novel low-bitrate neural speech codec based on parallel additive vector quantization (PAVQ). 
Unlike the mainstream residual vector quantization (RVQ) used in most neural speech codecs, where vector quantizers (VQs) are sequentially dependent, the PAVQ strategy adopted in PACodec aggregates parallel quantization results to optimize bitrate usage. 
Specifically, the PAVQ adopts a ``global–local–global'' (GLG) design: the global encoded features are quantized in parallel by multiple independent VQs, each attending to a local component of the representation, and their outputs are aggregated through addition to yield the final global quantization result for decoding. Experimental results show that PACodec, as each VQ focuses only on local information, supports smaller codebooks and reduces bitrate by 30\% compared with baselines at the same decoding quality, with only minor model complexity. Further analysis shows that, owing to the GLG framework of PAVQ, the proposed PACodec is disentanglement-friendly, and each independent VQ captures different aspects of speech, e.g., content, timbre, and acoustic details, suggesting potential for application to downstream tasks such as voice conversion.
\end{abstract}


\section{Introduction}
\renewcommand{\thefootnote}{}
\footnote{$^*$ Corresponding author. This work was funded by the National Nature Science Foundation of China under Grant 62301521.}
 \renewcommand{\thefootnote}{\arabic{footnote}}
\addtocounter{footnote}{-1}

Speech codecs compress speech signals to reduce the data required for representation while preserving acceptable decoding quality. 
They are a fundamental component of digital speech processing and are widely applied in speech communication, compression \cite{kim2025neural,o2023review}, and downstream tasks including speech synthesis \cite{shen2024naturalspeech,zhang2023speechgpt,zhang2023speak,achiam2023gpt,wang2023neural}, speech enhancement \cite{yang2024genhancer,yao2025gense}, and etc.
Bitrate is a key metric for evaluating speech codecs, as lower bitrates enable more efficient storage and transmission without sacrificing speech intelligibility or naturalness. 
Low-bitrate codecs are particularly valuable in bandwidth-constrained applications such as mobile communication, internet telephony, and large-scale speech processing systems.

Recently, with the advancement of deep learning, data-driven neural speech codecs have surpassed traditional ones \cite{valin2012opus,dietz2015overview}, delivering high decoded speech quality at substantially reduced bitrates. 
As a pioneer, SoundStream \cite{zeghidour2021soundstream} employs 1D convolutions and residual networks to directly encode speech waveforms, and it is the first to introduce generative adversarial networks (GANs) into the speech coding domain for better reconstructed speech quality. 
Subsequent neural speech codecs such as Encodec \cite{defossezhigh}, AudioDec \cite{wu2023audiodec}, DAC \cite{kumar2023high}, and HiFi-Codec \cite{yang2023hifi}, build on SoundStream and achieve further performance improvements.
However, directly modeling raw speech waveforms requires multiple downsampling and upsampling neural operations, leading to increased computational complexity. 
To address this, recent approaches shift from direct waveform modeling to the discretization of spectral features. 
For instance, APCodec \cite{ai2024apcodec} represents speech using amplitude and phase spectra as parametric features and employs a ConvNeXt v2 network \cite{woo2023convnext} for spectral coding, thereby significantly reducing model size. 
Furthermore, MDCTCodec \cite{jiang2024mdctcodec} leverages the modified discrete cosine transform (MDCT) spectrum, which is simpler and more compression-efficient. 
Compared with APCodec, which requires a dual-branch structure to model both amplitude and phase spectra, MDCTCodec adopts a single-branch design with much lower computational cost, thereby further streamlining codec design.

\begin{figure*}[t]
    \centering
    \includegraphics[width=\linewidth]{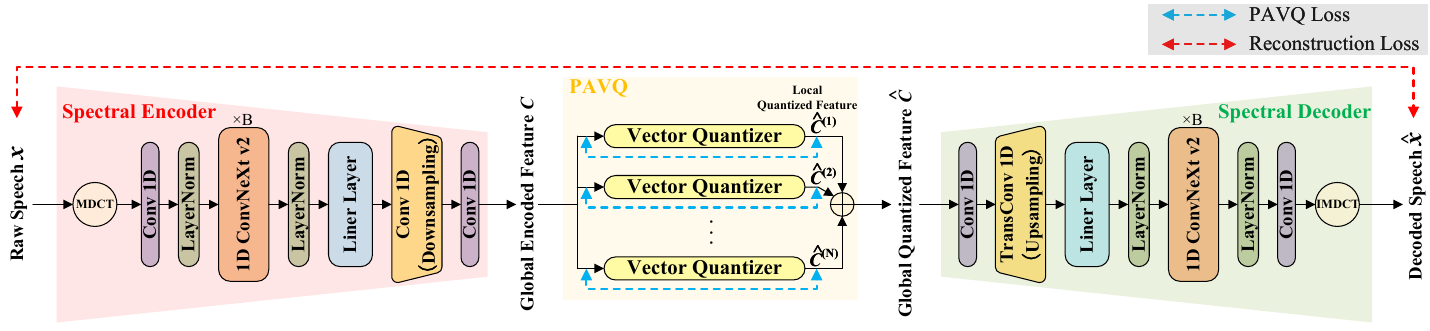}
    \caption{Overview of the proposed PACodec. Here, Conv 1D and TransConv 1D are 1D convolution and transposed convolution, respectively.}
    \label{fig:model}
\end{figure*}

Most neural speech codecs adopt residual vector quantization (RVQ) to discretize encoded features. 
In RVQ, sequentially dependent vector quantizers (VQs) iteratively refine the representation, but this interdependent structure makes further bitrate reduction difficult and restricts its applicability to tasks that require speech disentanglement.
To address these, HiFi-Codec \cite{yang2023hifi} extends RVQ with a grouped strategy (GRVQ), which splits encoded features by channel and applies parallel RVQs to each group. 
While this introduces partial parallelism, the reliance on RVQ still limits bitrate reduction, and channel-based grouping makes it difficult to interpret the role of each branch.
More recently, some researchers have shifted attention from vector quantization to finite scalar quantization (FSQ), as in SQCodec \cite{zhai2025one}, which adopts a single-quantizer design to avoid the complexity of multiple VQs. 
However, FSQ has been relatively underexplored, and to achieve competitive performance it requires an excessively large codebook, which hinders bitrate reduction.

To address the above limitations, we propose PACodec, a low-bitrate neural speech codec built on parallel additive vector quantization (PAVQ).
Unlike mainstream quantization strategies, the PAVQ used in PACodec employs a multi-branch parallel design with a ``global–local–global" (GLG) structure, where each branch contains a single VQ. 
All branches quantize the same global encoded features independently, each focusing on a local component of the representation, and their outputs are aggregated through addition to form the final global quantized result.
The PACodec is trained with the quantization loss of PAVQ and the speech reconstruction loss. 
Experimental results demonstrate that PACodec can operate with a codebook size as small as 128 by focusing each VQ on local information and, compared with baseline codecs, consistently achieves about 30\% bitrate reduction across both 16 kHz and 48 kHz datasets while preserving speech quality.
Further analysis shows that, owing to the GLG framework, each VQ in PAVQ captures different local speech information (e.g., content, timbre, acoustic details), indicating disentanglement potential and motivating future applications of PACodec to downstream tasks such as voice conversion (VC).

\section{Proposed Method}
\label{sec:method}

\subsection{Overview}

An overview of the proposed PACodec is shown in Fig. \ref{fig:model}. 
PACodec consists of a spectral encoder, a PAVQ, and a spectral decoder. 
The spectral encoder compresses the input speech into features, which are quantized by PAVQ and then decoded by the spectral decoder to reconstruct the waveform. 
During training, the PACodec adopts the quantization loss of PAVQ to minimize quantization errors and the reconstruction loss to narrow the gap between the decoded and raw speech.

\begin{table*}[t]
    \centering
    \caption{Experimental results of speech coding performance comparison. For PACodec, the bitrate is 1.4 kbps on LibriTTS and 4.2 kbps on VCTK. For baselines, (L) denotes low bitrate setting (1.5 kbps for LibriTTS and 4.5 kbps for VCTK), while (H) denotes high bitrate setting (2 kbps for LibriTTS and 6 kbps for VCTK). The \textbf{bold} and \underline{underline} numbers indicate optimal and sub-optimal results, respectively.}
    \label{tab:coding}
    \resizebox{\textwidth}{!}{
    \begin{tabular}{cc|cccc|cccc|c}
    \hline

    \hline
         \multirow{2}*{Model} & \multirow{2}*{Quantization} & \multicolumn{4}{c|}{LibriTTS (16 kHz)} & \multicolumn{4}{c|}{VCTK (48 kHz)} & \multirow{2}*{Param.$\downarrow$} \\ \cline{3-10}
         & & LSD (dB)$\downarrow$ & STOI$\uparrow$ & UTMOS$\uparrow$ & DNSMOS$\uparrow$ & LSD (dB)$\downarrow$ & STOI$\uparrow$ & UTMOS$\uparrow$ & DNSMOS$\uparrow$ & \\ \hline
         \textbf{PACodec} & PAVQ & 0.89 & \underline{0.93} & 3.80 & \underline{3.26} & \textbf{0.78} & \underline{0.89} & 3.93 & \textbf{3.17} & \textbf{6.7M}  \\ \hline
         \textbf{Encodec} (L) & RVQ & 0.97 & 0.89 & 3.30 & 3.17 & 0.91 & 0.84 & 3.69 & 3.12 & 17.6M  \\
         \textbf{AudioDec} (L) & RVQ & 0.96 & 0.71 & 2.95 & 3.09 & 0.86 & 0.79 & 3.61 & 3.12 & 24.4M  \\
         \textbf{DAC} (L) & RVQ & 0.90 & 0.92 & 3.65 & \underline{3.26} & 0.85 & 0.87 & 3.87 & 3.14 & 76.5M \\
         \textbf{APCodec} (L) & RVQ & 1.33 & 0.75 & 1.71 & 2.68 & 0.83 & 0.86 & 3.92 & 3.13 & 17.1M \\
         \textbf{MDCTCodec} (L) & RVQ & \underline{0.88} & \underline{0.93} & \underline{3.87} & \textbf{3.28} & 0.83 & 0.88 & 3.77 & 3.14 & \underline{6.8M}  \\ 
         \textbf{SQCodec} (L) & FSQ & 0.98 & 0.92 & \textbf{3.88} & 3.23 & - & - & - & - & 15.1M \\ \hline
         \textbf{Encodec} (H) & RVQ & 0.95 & 0.91 & 3.48 & 3.16 & 0.90 & 0.85 & 3.70 & 3.10 & 17.6M \\
         \textbf{AudioDec} (H) & RVQ & 0.95 & 0.72 & 3.08 & 3.09 & 0.85 & 0.80 & 3.68 & 3.11 & 24.4M \\
         \textbf{DAC} (H) & RVQ & 0.90 & \underline{0.93} & 3.64 & \underline{3.26} & 0.84 & \textbf{0.91} & \underline{3.94} & 3.14 & 76.5M \\
         \textbf{APCodec} (H) & RVQ & 1.29 & 0.74 & 1.55 & 2.51 & \underline{0.82} & 0.87 & \textbf{3.95} & \underline{3.15} & 17.1M \\
         \textbf{MDCTCodec} (H) & RVQ & \textbf{0.86} & \textbf{0.94} & \textbf{3.88} & 3.25 & 0.83 & \underline{0.89} & 3.84 & 3.13 & \underline{6.8M} \\ 
         \textbf{HiFi-Codec} (H) & GRVQ & 1.00 & 0.91 & 3.70 & 3.18 & 0.85 & 0.88 & 3.80 & 3.13 & 63.6M  \\
         \hline

         \hline
    \end{tabular}}
\end{table*}

\subsection{Spectral Encoder \& Decoder}

The spectral encoder compresses the input raw speech $\bm{x}\in \mathbb{R}^T$ with sampling rate $f_s$ (Hz) into features $\bm{C}\in \mathbb{R}^{L \times K}$, and the spectral decoder reconstructs the speech $\hat{\bm{x}}\in \mathbb{R}^T$ from the PAVQ-quantized features $\hat{\bm{C}}\in \mathbb{R}^{L \times K}$, where $T$ is the speech waveform length and $L$ and $K$ are the number of frames and channels, respectively. 
The ratio $R=T/L$ corresponds to the downsampling/upsampling rate of the spectral encoder/decoder. 

Inspired by \cite{jiang2024mdctcodec}, the spectral encoder and decoder use the highly compressed MDCT spectrum as the coding target. 
The structures of the spectral encoder and decoder are mirror-symmetric. 
In the spectral encoder, the input raw speech $\bm{x}$ is first transformed into the MDCT spectrum and then processed with 1D convolution and layer normalization. 
The resulting representations are then fed into $B$ cascaded 1D ConvNeXt v2 networks for deeper processing.
Each 1D ConvNeXt v2 network is composed of a stack of modules, including a 1D depthwise convolution, layer normalization, linear transformations, Gaussian error linear unit (GELU) activation, global response normalization (GRN), and a final linear projection, with residual connections applied to produce the output. 
The output of the last network is post-processed with layer normalization and a linear layer, then downsampled and reduced in dimension through two 1D convolutions to produce the encoded features $\bm{C}$ for subsequent discretization. 
The value of $R$ is jointly determined by the frame shift of the MDCT and the stride of the downsampling 1D convolutions.
The spectral decoder mirrors the encoder, replacing downsampling convolutions with upsampling transposed convolutions, and reconstructs the waveform $\hat{\bm{x}}$ via inverse MDCT (IMDCT).

\subsection{Parallel Additive Vector Quantization}
\label{sec: PAVQ}

The PAVQ quantizes the features $\bm{C}=[\bm{c}_1,\bm{c}_2,\dots,\bm{c}_L]^{\top}$ to produce the results $\hat{\bm{C}}=[\hat{\bm{c}}_1,\hat{\bm{c}}_2,\dots,\hat{\bm{c}}_L]^{\top}$. 
PAVQ is composed of $N$ parallel VQs under the GLG design, with each VQ quantizing a different local aspect of speech. 
This localized scheme facilitates the use of smaller codebooks for bitrate reduction. 
Accordingly, $\bm{C}$ and $\hat{\bm{C}}$ are referred to as the global encoded and quantized features, respectively. 
Specifically, each VQ is equipped with a trainable codebook $\mathbb{W}_n=\{\bm{w}_m^{(n)}\in\mathbb{R}^K \mid m=1,\dots,M^{(n)}\}$, where $n=1,\dots,N$ indexes the VQs and $M^{(n)}$ denotes the size of the $n$-th codebook. 
Each VQ takes the same global encoded feature $\bm{C}$ as input and selects quantization vectors from its codebook according to the minimum Euclidean distance principle. 
Taking the $n$-th VQ as an example, for the $l$-th encoded feature vector $\bm{c}_l \in \mathbb{R}^K$ from $\bm{C}$, where $l=1,\dots,L$, its quantized vector $\hat{\bm{c}}_l^{(n)}$ and corresponding token $d_l^{(n)}$ are obtained as
\begin{equation}
    \hat{\bm{c}}_l^{(n)},d_l^{(n)}=\arg\min_{\bm{w}_m^{(n)},m} \| \bm{c}_l - \bm{w}_m^{(n)} \|_2,
\end{equation}
where $\hat{\bm{c}}_l^{(n)}\in\mathbb R^K$ and $d_l^{(n)}\in \{1,\dots,M^{(n)} \}$.
Therefore, the local quantized feature produced by the $n$-th VQ is $\hat{\bm{C}}^{(n)}=[\hat{\bm{c}}^{(n)}_1,\dots,\hat{\bm{c}}^{(n)}_{L}]^{\top}\in \mathbb{R}^{L \times K}$. 
Finally, the local quantized features produced by all VQs are aggregated through addition to form the global quantized feature $\bm{\hat{C}}$, i.e.,
\begin{equation}
\label{equ: add}
    \hat{\bm{C}}=\sum_{n=1}^N \hat{\bm{C}}^{(n)}.
\end{equation}
The tokens produced by PAVQ are $\bm{d}^{(1)},\dots,\bm{d}^{(N)}$, where $\bm{d}^{(n)}=[d^{(n)}_1,\dots,d^{(n)}_L]^{\top}, n=1,\dots,N$ and the bitrate is calculated as
\begin{equation}
    Bitrate=\frac{f_s}{R} \cdot \log_2\prod_{n=1}^N  M^{(n)} \quad(bps).
\end{equation}

Unlike RVQ, the proposed PAVQ allows each VQ to capture relatively independent local information, eliminating the need for large codebooks and thereby reducing bitrate. 
Unlike GRVQ, it avoids residual structures and channel-wise grouping, yielding a simpler design and clearer interpretation of each VQ.

\subsection{Training Criteria}

During training, PACodec is jointly optimized at both the quantization and reconstruction levels. 
At the quantization level, we define the PAVQ loss to optimize the trainable codebooks and reduce quantization errors. 
It is formulated as the sum of the mean squared errors (MSE) between the inputs and outputs of all VQs across the branches, i.e., 
\begin{equation}
\mathcal{L}_{\mathrm{PAVQ}} 
= \sum_{n=1}^N \mathbb{E}_{(\hat{\bm{C}}^{(n)},\,\bm{C})} 
\left\| \hat{\bm{C}}^{(n)} - \bm{C} \right\|_{F}^{2},
\end{equation}
where $\left\| \cdot \right\|_{F}$ is the Frobenius norm.

At the reconstruction level, inspired by \cite{jiang2024mdctcodec}, we define the loss between the decoded speech $\hat{\bm{x}}$ and raw speech $\bm{x}$, consisting of adversarial and spectral reconstruction terms, i.e.,
\begin{equation}
\mathcal{L}_{\mathrm{recon}}=\mathcal{L}_{\mathrm{adv}}(\bm{\hat{x}},\bm{x})+\mathcal{L}_{\mathrm{spec}}(\bm{\hat{x}},\bm{x}).
\end{equation}
For adversarial loss $\mathcal{L}_{\mathrm{adv}}$, we adopt a multi-resolution MDCT-based discriminator, which applies 2D convolutions to judge the MDCT spectra of raw and decoded speech under different MDCT configurations.
For spectral reconstruction loss $\mathcal{L}_{\mathrm{spec}}$, we measure the difference between decoded and raw speech in both MDCT spectrum and mel-spectrogram domains, using MSE for MDCT spectra and a combination of MSE and mean absolute error (MAE) for mel-spectrograms. 
This encourages the decoded speech to perceptually align with raw speech.

Finally, we combine the two losses to jointly train PACodec, i.e.,
\begin{equation}
\mathcal{L}=\mathcal{L}_{\mathrm{PAVQ}}+\mathcal{L}_{\mathrm{recon}}.
\end{equation}

\section{Experiments and Results}
\subsection{Experimental Setting}

To evaluate our proposed PACodec across different sampling rates, we conducted experiments on the LibriTTS \cite{zen2019libritts} and VCTK \cite{yamagishi2019cstr} datasets. 
The 16 kHz downsampled LibriTTS dataset, containing about 585 hours of speech with 354,780 training utterances from 2,311 speakers and 4,837 test utterances from 39 unseen speakers, was used in our experiments.
The VCTK dataset contains about 43 hours of 48 kHz speech. 
We used 40,936 utterances from 100 speakers for training, while the remaining 2,937 utterances from 8 unseen speakers were used for testing.

In our experiments\footnote{Speech samples are available at: \url{https://anonymity225.github.io/PACodec/}.}, the MDCT in PACodec used a frame length of 80, with both the frame shift and the number of frequency bins set to 40.
The spectral encoder and decoder each used 8 1D ConvNeXt v2 blocks (i.e., $B=8$).
All 1D convolutions, except the last one, produced 256 output channels, while the last 1D convolution output 32 channels (i.e., $K=32$) in the encoder and 40 channels in the decoder. 
All convolution kernels had a size of 7.
The upsampling and downsampling factors in the encoding and decoding stages were both set to 320 (i.e., $R=320$).
The PAVQ used 4 VQs (i.e., $N=4$), each with a codebook size of 128 (i.e., $M^{(*)}=128$). 
For training, we used the AdamW optimizer ($\beta_1=0.8,\ \beta_2=0.99$) with an initial learning rate of 0.0002 decayed by 0.999 per epoch, for 500 epochs in total.

\subsection{Baselines}

We compared the proposed PACodec with several advanced baseline neural speech codecs, i.e., Encodec \cite{defossezhigh}, AudioDec \cite{wu2023audiodec}, DAC \cite{kumar2023high}, HiFi-Codec \cite{yang2023hifi}, APCodec \cite{ai2024apcodec}, MDCTCodec \cite{jiang2024mdctcodec}, and SQCodec \cite{zhai2025one}.
To ensure a fair comparison and highlight the performance of PACodec, at LibriTTS dataset (16 kHz) we compared PACodec at 1.4 kbps with baseline codecs at 1.5 kbps and 2 kbps. 
At VCTK dataset (48 kHz), we compared PACodec at 4.2 kbps with baseline codecs at 4.5 kbps and 6 kbps. 
SQCodec was only included at 1.5 kbps for 16 kHz, as no other settings were provided in its official release.
HiFi-Codec, due to the characteristics of GRVQ, cannot operate at 1.5 kbps for 16 kHz or 4.5 kbps for 48 kHz, and thus these results were excluded.
All baseline codecs were re-trained using their official implementations.

\begin{figure}[t]
  \centering
  \includegraphics[width=\linewidth]{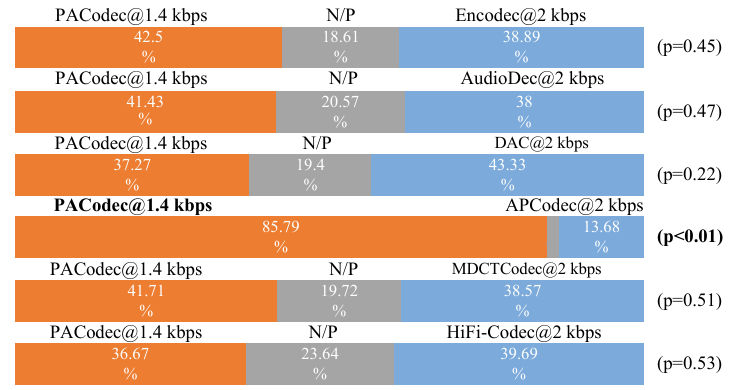}
  \caption{
  Average preference scores (\%) of ABX tests comparing PACodec at 1.4 kbps and baselines at 2 kbps on LibriTTS (16 kHz). N/P denotes “no preference”, and $p$ is the paired $t$-test $p$-value.}
  \label{fig: ABX}
\end{figure}

\subsection{Speech Coding Performance Comparison}

To compare the coding performance of PACodec with baselines, we used both objective and subjective evaluations.
For objective evaluation, we adopted four measures: log spectral distance (LSD), short-time objective intelligibility (STOI), UTMOS \cite{saeki2022utmos}, and DNSMOS \cite{reddy2022dnsmos}. 
The LSD and STOI reflect spectral quality and intelligibility, respectively, while UTMOS and DNSMOS are non-intrusive metrics that provide perceptual scores of speech.
In addition, we also used the number of model parameters (Param.) to evaluate the model complexity. 
For subjective evaluation, we conducted ABX paired preference tests on Amazon Mechanical Turk, where 40 native English-speaking listeners compared 10 utterance pairs per method to judge quality or no preference. 
We reported average preference scores and $t$-test $p$-values for statistical significance.

The objective results are summarized in Table \ref{tab:coding}. On the LibriTTS dataset, PACodec at 1.4 kbps achieved performance comparable to 2 kbps baselines, showing that it can maintain similar quality at a much lower bitrate. 
On the VCTK dataset, the advantage of PACodec was more pronounced: at 4.2 kbps, it not only outperformed the 4.5 kbps baselines but in some metrics even surpassed the 6 kbps codecs, thereby saving 1.8 kbps (i.e., 30\% bitrate saving) while delivering equal or better quality. 
Since the objective metrics on LibriTTS were less discriminative than those on VCTK, we further conducted subjective ABX preference listening tests as a supplement. 
As shown in Fig. \ref{fig: ABX}, PACodec at 1.4 kbps delivered perceptual quality on par with 2 kbps baselines, achieving a 600 bps bitrate saving (i.e, 30\% bitrate saving) without significantly reducing perceived naturalness ($p>0.01$).


In addition, as shown in Table \ref{tab:coding}, PACodec has the smallest number of parameters among all codecs, with only 6.7M. This further demonstrates its lower complexity and higher practicality. Overall, PACodec not only achieves strong coding performance at lower bitrates but also maintains lightweight model complexity, making it well suited for real-world applications.

\subsection{Disentanglement Potential Analysis}

\begin{table}[t]
    \centering
    \caption{Experimental results of PACodec and its ablated variants on the VCTK dataset, where \textit{italics} indicate severely degraded metrics.}
    \label{tab:jieou}
    \resizebox{\columnwidth}{!}{
    \begin{tabular}{rcccccc}
    \hline

    \hline
         & WER$\downarrow$ & CER$\downarrow$ & STOI$\uparrow$ & F0-RMSE (cent)$\downarrow$ & SS$\uparrow$ & MCD (dB)$\downarrow$ \\ \hline
         \textbf{PACodec} & 0.03 & 0.01 & 0.89 & 30.9 & 0.81 & 1.53 \\
         \textbf{-VQ$_1$} & \textit{0.75} & \textit{0.59} & \textit{0.46} & 203.1 & \textit{0.22} & \textit{10.09} \\
         \textbf{-VQ$_2$} & 0.31 & 0.20 & 0.83 & 30.6& 0.51 & 2.44  \\
         \textbf{-VQ$_3$} & 0.06 & 0.03 & 0.86 & 36.1& 0.60  & 2.20 \\
         \textbf{-VQ$_4$} & 0.19 & 0.11 & 0.69 & \textit{1114.0} & \textit{0.39}  & 5.08 \\ 
         \hline

         \hline
    \end{tabular}}
\end{table}

As introduced in Section \ref{sec: PAVQ}, PACodec adopts the GLG framework of PAVQ, which allows each VQ to attend to local speech attributes. 
We therefore conduct experiments to assess its disentanglement potential by analyzing the types of information contained in these local representations on the VCTK dataset (48 kHz). 
To examine the role of each VQ branch, we conducted ablation experiments by removing its quantized output and reconstructing speech with the remaining branches. 
For example, when ablating the $n'$-th VQ (-VQ$_{n'}$), Equation \ref{equ: add} becomes $\hat{\bm{C}}=\sum_{n=1,n\ne n'}^N \hat{\bm{C}}^{(n)}$. 
This highlights the advantage of the addition operation in PAVQ over the concatenation operation in GRVQ, as it facilitates partial speech reconstruction for analysis.


We adopted three categories of evaluation metrics. 
Content-related metrics include word error rate (WER), character error rate (CER), and short-time objective intelligibility (STOI). 
Timbre-related metrics include root mean square error of F0 (F0-RMSE) and speaker similarity (SS). 
Finally, perceptual acoustic detail is evaluated using mel-cepstral distortion (MCD). 

As shown in Table \ref{tab:jieou}, removing VQ$_1$ led to significant degradation across all metrics, especially WER, CER, and STOI, suggesting that VQ$_1$ mainly quantizes text-content information and is critical for speech intelligibility and comprehensibility, whose absence also affects other metrics. 
When VQ$_4$ was removed, content-related metrics declined only slightly compared with the removal of VQ$_1$, but timbre-related metrics, especially F0-RMSE, deteriorated severely, indicating that VQ$_4$ mainly quantizes pitch- and timbre-related information.
In contrast, removing VQ$_2$ or VQ$_3$ led to moderate degradation across the metrics but without severe declines, indicating that text content and timbre information were largely preserved. 
This suggests that VQ$_2$ and VQ$_3$ primarily quantize acoustic details to enhance the overall quality of decoded speech, with VQ$_2$ capturing richer acoustic detail than VQ$_3$.
These results confirm that, owing to the GLG framework, different PAVQ branches capture complementary local information such as content, timbre, and acoustic details, demonstrating disentanglement potential. 
This motivates future work on applying PACodec to disentanglement-based tasks such as VC.

\begin{table}[t]
    \centering
    \caption{Experimental results of PACodec and its lower-bitrate variant on the VCTK dataset.}
    \label{tab:3VQ}
    \resizebox{\columnwidth}{!}{
    \begin{tabular}{ccccc}
    \hline

    \hline
         & LSD (dB)$\downarrow$ & STOI$\uparrow$ & UTMOS$\uparrow$ & DNSMOS$\uparrow$ \\ \hline
         \textbf{PACodec@4.2 kbps} & 0.78 & 0.89 & 3.93 & 3.17  \\
         \textbf{PACodec@3.15 kbps} & 0.79 & 0.87 & 3.89 &  3.16 \\ 
         \hline

         \hline
    \end{tabular}}
\end{table}

\subsection{VQ Number Sensitivity Analysis}

Unlike RVQ and GRVQ, where VQs are tightly coupled through residual dependencies and the codec performance is highly sensitive to the number of VQs, the PAVQ used in our proposed PACodec operates with independent branches, which may reduce its sensitivity to VQ quantity.
Therefore, we attempted to ablate the number of PAVQ branches (reducing $N$ from 4 to 3) to evaluate PACodec’s performance at lower bitrates (from 4.2 kbps to 3.15 kbps), and the experimental results on the VCTK dataset (48 kHz) are shown in Table \ref{tab:3VQ}.
These results indicate that PACodec’s PAVQ is relatively insensitive to the number of VQs, demonstrating robustness under varying configurations and suggesting strong potential for further bitrate reduction without significant performance degradation.

\section{Conclusion}

This paper presents PACodec, a novel low-bitrate neural speech codec built on a PAVQ strategy. 
Unlike RVQ and GRVQ, PAVQ adopts a GLG design in which global encoded features are quantized by multiple independent branches, each capturing local information, and then aggregated through addition to yield the final global representation. 
Experimental results show that PACodec, by enabling smaller codebooks, achieves about 30\% bitrate reduction compared with baselines at the same decoding quality, with only minor model complexity. 
Further analysis confirms that PACodec is disentanglement-friendly, as different VQs capture complementary aspects of speech such as content, timbre, and acoustic details. 
In future work, we plan to explore PACodec’s applications in disentanglement-based downstream tasks, such as VC.

\printbibliography

\end{document}